\documentclass[aps,nofootinbib,showpacs,preprintnumbers]{revtex4}
\usepackage{graphicx}
\usepackage{amsfonts}
\usepackage{amssymb}

\begin{document}

\preprint{OU-HET 534}\preprint{DTP-MSU/05-08}

\preprint{hep-th/0506216}

\title{Intersecting Non-extreme $p$-Branes and Linear Dilaton Background}

\author{Chiang-Mei Chen} \email{cmchen@phy.ncu.edu.tw}
\affiliation{Department of Physics, National Central University,
Chungli 320, Taiwan}

\author{Dmitri V. Gal'tsov} \email{galtsov@mail.phys.msu.su}
\affiliation{Department of Theoretical Physics, Moscow State
University, 119899, Moscow, Russia}

\author{Nobuyoshi Ohta} \email{ohta@phys.sci.osaka-u.ac.jp}
\affiliation{Department of Physics, Osaka University, Toyonaka,
Osaka 560-0043, Japan}

\date{\today}

%\maketitle

\begin{abstract}
We construct the general static solution to the supergravity
action containing gravity, the dilaton and a set of antisymmetric
forms describing the intersecting branes delocalized in the
relative transverse dimensions. The solution is obtained by
reducing the system to a set of separate Liouville equations (the
intersection rules implying the separability); it contains the
maximal number of free parameters corresponding to the rank of the
differential equations. Imposing the requirement of the absence of
naked singularities, we show that the general configurations are
restricted to two and only two classes: the usual asymptotically
flat intersecting branes, and the intersecting branes some of
which are asymptotically flat and some approach the linear dilaton
background at infinity. In both cases the configurations are
black. These are supposed to be relevant for the description of
the thermal phase of the QFT's in the corresponding Domain-Wall/QFT duality.
\end{abstract}


\maketitle
%%%%%%%%%%%%%%%%%%%%%%%%%%%%%%%%%%%%%%%%%%%%%%%%%%%%%%%%%%%%%%%%%
\section{Introduction}
%%%%%%%%%%%%%%%%%%%%%%%%%%%%%%%%%%%%%%%%%%%%%%%%%%%%%%%%%%%%%%%%%
Recently it has been shown \cite{Clement:2004ii} that the general
supergravity solution describing a charged  $p$-brane without
naked singularities can be either the standard black
asymptotically flat (AF) $p$-brane, or the black brane which
asymptotically approaches the linear dilaton background (LDB). The
LDB  is known to be the near-horizon limit of the extremal
dilatonic branes; it is relevant for the description of
(non-conformal) quantum field theory (QFT) in the Domain-Wall/QFT
correspondence \cite{Boonstra:1998mp, Behrndt:1999mk,Cai:1999xg}
and (in the case of NS5 branes) little string theories (LST)
\cite{Aharony:1998ub, Aharony:1999ks}. According to the standard
reasoning, the same configuration endowed with an event horizon
should describe the thermal version of the same QFT (LST). The
asymptotically LDB black branes obtained must thus describe the
class of quantum field theories in the thermal phase, or, in the
particular case of NS5-branes, the thermodynamics of little string
theory. This was extensively studied recently in the case of
one-brane solutions \cite{HaOb,BeRo,BoRo}, so it is interesting to
present a more general intersecting brane framework.

The simplest example of the LDB endowed with an event horizon is
the four-dimensional charged dilatonic black hole with the LDB
asymptotics \cite{Chan:1995fr, Clement:2002mb, Clement:2004yr,Cai:2004iy}.
This configuration was also identified as ``the horizon plus
throat'' geometry \cite{Giddings:1992kn} arising  in the
near-horizon limit of the near extremal dilatonic black hole
\cite{Gibbons:1987ps, Garfinkle:1990qj}. The near-horizon limit of
the BPS dilatonic black hole is the LDB itself, and in the
theories admitting the 1/2 BPS branes the  limiting configuration
preserves the half of SUSY of the initial theory. The near horizon
limit of {\em black} dilatonic black hole \cite{Clement:2002mb}
keeps the memory about the event horizon and turns out to be a
non-supersymmetric configuration whose BPS limit is the LDB. The
brane generalization of this construction has interesting
particular cases in ten dimensions~\cite{Clement:2004ii}.

On the other hand, it has been also known that a more general
class of solutions consisting of various kinds of black branes can
be constructed within the same framework
\cite{Ohta:1997wp, Aref'eva:1997nz, Ohta:1997gw, Ohta:1997wd,
Ivashchuk:1998jg,Ohta:2003rr, Miao:2004bn, Rama:2005bd,Bai:2005jr}
(for the time-dependent branes see \cite{Chen:2002yq, Ohta:2003uw,
Ohta:2003zh}). It is then natural to ask if the above results can
be extended to such general solutions. The purpose of this paper
is to generalize the construction to the case of intersecting
branes (see for review \cite{Ga97, Lu98, Sm02}). We show that the
solutions are restricted to either the asymptotically flat black
branes or asymptotically LDB ones and their mixed system if we
impose the condition that there are no naked singularities. This
generalization is interesting in that it opens the possibility of
extending the Domain-Wall/QFT correspondence to more general
configurations.

This paper is organized as follows. In the next section, we start
with the action for the $D$-dimensional gravity coupled to the
dilaton and an arbitrary number of form fields of various ranks.
We summarize the field equations, metric ansatz and the background
for forms. We then derive the most general static solutions in the
theory requiring the usual intersection rules for the branes. The
obtained solutions have a large number of parameters. In Sec.~III,
we fix some of them by the requirement that the solutions do not
have naked singularities, and show that the resulting solutions
consist of either AF black branes or the black branes which
asymptotically approach the LDB. In Sec.~IV, we transform the
solutions to the more familiar coordinates, and show that these
reduce to the known solutions of intersecting AF branes and/or
branes in LDB. Of course, the latter can be obtained from the
known intersecting AF branes by taking the near-horizon limit, but
a naive application of this rule produces the solutions consisting
only of intersecting branes with the LDB asymptotics; our
additional solutions of mixed type of AF and asymptotically LDB
black branes can not be simply obtained in such a limit. Not only
this, our results show that these are {\em the only} solutions
which can be obtained by imposing the requirement that the
space-time is free of naked singularities.

%%%%%%%%%%%%%%%%%%%%%%%%%%%%%%%%%%%%%%%%%%%%%%%%%%%%%%%%%%%%%%%%%%%%%%
\section{Intersecting Non-extreme $p$-branes}
%%%%%%%%%%%%%%%%%%%%%%%%%%%%%%%%%%%%%%%%%%%%%%%%%%%%%%%%%%%%%%%%%%%%%%
We consider the action describing gravity coupled to a dilaton
$\phi$ and $m$ different $n_A$-form fields in $D$ dimensions
\begin{equation}
S = \int d^D x \sqrt{-g} \left( R - \frac12 \partial_\mu \phi
\partial^\mu \phi - \sum_{A=1}^m \frac1{2\, n_A!} \, {\rm e}^{a_A
\phi} \, F_{[n_A]}^2 \right).
\end{equation}
All form fields are coupled to the unique dilaton with individual
coupling constants $a_A$. This action is the simplest one to
describe intersecting supergravity $p$-branes, it incorporates
various (truncated) supergravity models for different choice of
the parameters $D,\, n_A,$ and $a_A$. The corresponding equations
of motion read
\begin{eqnarray}
R_{\mu\nu} - \frac12 \partial_\mu \phi \partial_\nu \phi -
\sum_{A=1}^m \frac{{\rm e}^{a_A \phi}}{2(n_A-1)!} \left[
(F_{[n_A]}^2)_{\mu\nu} - \frac{n_A-1}{n_A(D-2)} F_{[n_A]}^2 \,
g_{\mu\nu} \right] &=& 0, \label{Ein}
\\
\partial_\mu \left( \sqrt{-g} \, {\rm e}^{a_A \phi} \,
F^{\mu\nu_2\cdots \nu_{n_A}} \right) &=& 0, \label{form}
\\
\frac1{\sqrt{-g}}\, \partial_\mu \left( \sqrt{-g} \partial^\mu
\phi \right) - \sum_{A=1}^m \frac{a_A}{2 \, n_A!} {\rm e}^{a_A
\phi} F_{[n_A]}^2 &=& 0.
\label{dil}
\end{eqnarray}
In addition, each form field satisfies the Bianchi identity
\begin{equation}
\partial_{[\mu} F_{\nu_1\nu_2 \cdots \nu_{n_A}]} = 0.
\end{equation}
The notation $(F_{[n_A]}^2)_{\mu\nu}$ is defined as
\begin{equation}
(F_{[n_A]}^2)_{\mu\nu} := F_{\mu\alpha_2\cdots \alpha_{n_A}}
F_\nu{}^{\alpha_2\cdots \alpha_{n_A}}.
\end{equation}

The most general ansatz for the metric describing the set of
intersecting non-extremal branes is
%(index $\rho$ is reserved for labeling $x^\rho$)
\begin{equation}\label{metric}
ds^2 = - {\rm e}^{2B} dt^2 + \sum_{\rho=1}^p {\rm e}^{2 C_\rho}
dx_\rho^2 + {\rm e}^{2A} \, dr^2 + {\rm e}^{2D} \,
d\Sigma_{k,\sigma}^2,
\end{equation}
where the overall transverse space is described by
\begin{equation}
d\Sigma_{k,\sigma}^2 = \bar g_{ab} dz^a dz^b = \left\{
\begin{array}{ll}
 d \psi^2 + \sin^2\psi \, d\Omega_{k-1}^2, \qquad & \sigma=+1,\\
 d \psi^2 + \psi^2 \, d\Omega_{k-1}^2, \qquad & \sigma=0,\\
 d \psi^2 + \sinh^2\psi \, d\Omega_{k-1}^2, \qquad & \sigma=-1,
 \end{array} \right.
\label{gmetric}
\end{equation}
satisfying
\begin{equation}
\bar R_{ab} = \sigma (k - 1) \bar g_{ab}.
\end{equation}
The choice of $\sigma$ corresponds to  different symmetries of the
overall transverse space, namely, $SO(k)$ for $\sigma=1$, $E(k)$
for for $\sigma=0$, and $SO(1, k-1)$ for $\sigma=-1$. Two latter
cases correspond to topologically non-trivial solutions which are
mostly known in the (multidimensional) black hole case. In what
follows we will derive the general solution valid for all
$\sigma$, but later on we restrict to the case $\sigma=1$. For a
discussion of the topological solutions (in the single brane case)
see, e.g., \cite{Gr01,Gal'tsov:2004kn}. The total number of
dimensions occupied by the brane world-volumes is $p$. Each brane
of the set is specified by the particular form field labeled by
$n_A$.

The Ricci-tensor for this metric has the following non-vanishing
components:
\begin{eqnarray}
R_{tt} &=& {\rm e}^{2B-2A} (B'' + B' \Lambda'),
\label{Rtt} \\
R_{x_\rho x_\rho} &=& - {\rm e}^{2C_\rho-2A} (C_\rho'' + C_\rho'
\Lambda'),
\label{Rxx} \\
R_{rr} &=& - \Lambda'' - A'' + A' \Lambda' + A'^2 - B'^2 -
\sum_{\rho=1}^p C_\rho'^2 - k D'^2,
\label{Rrr} \\
R_{ab} &=&  \left[ - {\rm e}^{2D-2A} (D'' + D' \Lambda') + \sigma
(k-1) \right] \, \bar g_{ab}, \label{Rab}
\end{eqnarray}
where
\begin{equation}
\Lambda = - A + B + \sum_{\rho=1}^p C_\rho + k D.
\label{gauge1}
\end{equation}

We consider both the electrically and magnetically charges branes.
In the electric case  each form field is given by
\begin{equation}
F_{[n_A]} = f_A^{\rm elec} \, dt \wedge dx_{\rho_3} \wedge \cdots
\wedge dx_{\rho_{n_A}} \wedge dr,
\end{equation}
which satisfies the Bianchi identity automatically. It supports
the $p_A(=n_A-2)$-brane in the set of intersecting branes. The
equation of motion for the form fields is solved as
\begin{equation}
f_A^{\rm elec} = b_A \exp\left( A + B - \sum_{\rho=1}^p C_\rho -
kD + 2 \sum_{\rho=1}^p C_{\rho} \delta^{\rho}_A - a_A \phi\right),
\label{elb}
\end{equation}
where $\delta^{\rho}_A = 1$ for $x^\rho$ belonging to the
world-volume of the $p_A$-brane.

The form fields for the magnetic branes read
\begin{equation}
F_{[n_A]} = b_A dx_{\rho_1} \wedge \cdots \wedge dx_{\rho_{n_A-k}}
\wedge {\rm vol}(\Sigma_{k,\sigma}),
\end{equation}
in which case the set $x^{\rho_i}$ does not belong to the
world-volume of the $p_A(=D-n_A-2)$-brane. It is easy to see that
the field equations for these form fields are satisfied indeed.

To solve the Einstein equations  we change the independent
variable as
\begin{equation}\label{tau}
d\tau = {\rm e}^{-\Lambda} \, dr,
\end{equation}
and present the radial part of the metric as
\begin{equation}
{\rm e}^{2A} dr^2 = {\rm e}^{2 {\cal A}} d\tau^2, \qquad  {\cal A}
= \Lambda + A.
\label{gauge2}
\end{equation}
The Einstein and dilaton equations to be solved are (with
derivatives with respect to $\tau$)
\begin{eqnarray}\label{EqBs}
B'' &=& \sum_{A=1}^m \frac{b_A^2 (D-p_A-3)}{2(D-2)} {\rm e}^{G_A},
\\
C_{\rho}'' &=&  \sum_{A=1}^m \frac{b_A^2
\delta_A^{(\rho)}}{2(D-2)}{\rm e}^{G_A},
\\
D'' &=& - \sum_{A=1}^m \frac{b_A^2 (p_A+1)}{2(D-2)} {\rm e}^{G_A}
+ \sigma (k-1) {\rm e}^{2{\cal A}-2D},
\\
\phi'' &=& - \sum_{A=1}^m \frac{\epsilon_A a_A b_A^2}2 {\rm
e}^{G_A},
\label{Eqphis}
\end{eqnarray}
where
\begin{equation}
\label{GA} G_A = - \epsilon_A a_A \phi + 2B + 2 \sum_{\rho=1}^p
C_{\rho} \delta^{\rho}_A, \label{ga}
\end{equation}
and we  have defined
\begin{eqnarray}
\delta_A^{(\rho)} = \left\{ \begin{array}{l}
D-p_A-3 \\
-(p_A+1)
\end{array}
\right. \hspace{5mm} {\rm for} \hspace{3mm} \left\{
\begin{array}{l}
x_\rho \mbox{ belonging to $p_A$-brane} \\
{\rm otherwise}
\end{array}.
\right. \label{8}
\end{eqnarray}
Note that this can be also written as
\begin{eqnarray}
\delta_A^{(\rho)} =(D-2) \delta_A^\rho -(p_A+1). \label{delta}
\end{eqnarray}

 The $rr$ component of Einstein equations gives
\begin{equation}\label{rr}
{\cal A}'' - {\cal A}'^2 + B'^2 + \sum_{\rho=1}^p C_\rho'^2 + k
D'^2 = - \frac12 \phi'^2 + \sum_{A=1}^m
\frac{b_A^2(D-p_A-3)}{2(D-2)} {\rm e}^{G_A}.
\end{equation}
It is straightforward to check that the following combination
separates:
\begin{eqnarray}\label{AD}
({\cal A} - D)''
&=& \sum_{A=1}^m \frac{b_A^2}{2(D-2)} \left[
D-p_A-3 + \sum_{\rho=1}^p \delta^{(\rho)}_A - (k-1)(p_A+1) \right]
{\rm e}^{G_A} + \sigma (k-1)^2 {\rm e}^{2({\cal A}-D)}
\nonumber\\
&=& \sigma (k-1)^2 {\rm e}^{2({\cal A}-D)},
\end{eqnarray}
where we have taken into account the relation
\begin{equation}
D-p_A-3 + \sum_{\rho=1}^p \delta^{(\rho)}_A = (k-1)(p_A+1),
\end{equation}
which follows from (\ref{8}).
The general solution of Eq.~(\ref{AD}) is
\begin{equation}
H := 2 ({\cal A} - D) = \left\{ \begin{array}{ll}
 \ln \left( \frac{\beta^2}{4(k-1)^2} \right) - \ln \left[
 \sinh^2\left( \frac{\beta}2(\tau-\tau_0) \right) \right] \qquad &
 \sigma=1, \\
 \beta (\tau - \tau_0), & \sigma=0, \\
 \ln \left( \frac{\beta^2}{4(k-1)^2} \right) - \ln \left[
 \cosh^2\left( \frac{\beta}2(\tau-\tau_0) \right) \right] \qquad &
 \sigma=-1.
\end{array}
\right.
\label{solh}
\end{equation}

Differentiating Eq.~(\ref{GA}) twice and
%, we obtain
%\begin{equation}
%G_A'' = - \epsilon_A a_A \phi'' + 2 B'' + 2 \sum_{\rho=1}^p
%C_{\rho}'' \delta^{\rho}_A.
%\end{equation}
substituting (\ref{EqBs}) -- (\ref{Eqphis}) into this, we arrive
at the following coupled system of equations for the set of $G_A$:
\begin{equation}
G_A'' = \sum_B \left\{ \frac{\epsilon_A a_A \epsilon_B a_B}2 +
\frac{D-p_A-3}{D-2} + \sum_{\rho=1}^p \frac{\delta_B^{(\rho)}
\delta_A^\rho}{D-2} \right\} b_B^2 \, {\rm e}^{G_B},
\end{equation}
where the relation (\ref{delta}) is understood. The matrix on the
right had side of this equation is non-diagonal generally. However
if we impose the  standard {\it intersection rule} for
$A \ne B$~\cite{Ohta:1997gw}:
\begin{equation}\label{intersection}
\sum_{\rho=1}^p \delta^{\rho}_A \delta^{\rho}_B \equiv \bar p =
\frac{(p_A+1)(p_B+1)}{D-2} - 1 - \frac{\epsilon_A a_A \epsilon_B a_B}2,
\label{int}
\end{equation}
where $\bar p$ relates to the brane on which $p_A$- and
$p_B$-branes intersect, the non-diagonal part vanishes. In this
case the system reduces to the set of separate Liouville
equations, and it is integrable. Therefore, the intersection rule
is a sufficient condition for integrability. It is an interesting
question whether the intersection rules is a {\it necessary}
condition for integrability, see the corresponding discussion  in
\cite{GaRy98}.

For the diagonal terms $A=B$, the expression in the parenthesis
reduces to
\begin{equation}
\{ \cdots \} = \frac{\Delta_A}2, \quad {\rm where} \quad \Delta_A
= a_A^2 + \frac{2(p_A+1)(D-p_A-3)}{D-2}.
\end{equation}
With this assumption, one obtains the set of the decoupled
Liouville equations for $G_A$:
\begin{equation}
G_A'' = \frac{\Delta_A b_A^2}2 {\rm e}^{G_A},
\end{equation}
which leads to the solution
\begin{equation}
G_A = \ln \left( \frac{\alpha_A^2}{\Delta_A b_A^2} \right) - \ln
\left[ \sinh^2\left( \frac{\alpha_A}2(\tau-\tau_A) \right)\right].
\label{ga1}
\end{equation}
Using this result, we can immediately integrate Eqs.
(\ref{EqBs}-\ref{Eqphis}) to obtain
\begin{eqnarray}
B &=& \sum_{A=1}^m \frac{D-p_A-3}{(D-2)\Delta_A} \left(
G_A + g_A^{(1)} \tau + g_A^{(0)} \right) + B^{(1)} \tau + B^{(0)},
\label{B} \\
C_{\rho} &=& \sum_{A=1}^m \frac{\delta_A^{(\rho)}}{(D-2) \Delta_A}
\left( G_A + g_A^{(1)} \tau +
g_A^{(0)} \right) + C_{\rho}^{(1)} \tau + C_{\rho}^{(0)},
\label{C} \\
\phi &=& - \sum_{A=1}^m \frac{\epsilon_A a_A}{\Delta_A} \left( G_A
+ g_A^{(1)} \tau + g_A^{(0)} \right) + \phi^{(1)} \tau +
\phi^{(0)}. \label{phi}
\end{eqnarray}
It follows from Eq.~(\ref{GA}) that the constants are connected by
the following $2m$ relations:
\begin{equation}\label{ConBs}
g_A^{(0,1)} + 2 B^{(0,1)} + 2 \sum_{\rho=1}^p C_{\rho}^{(0,1)}
\delta^{\rho}_A - \epsilon_A a_A \phi^{(0,1)} = 0,
\end{equation}
%The sum of the functions $C_{\rho}$ is
%\begin{equation}
%\sum_{\rho=1}^p C_{\rho} = \sum_{A=1}^m \frac{k (p_A+1) -
%(D -2)}{(D-2) \Delta_A} \left( G_A + g_A^{(1)} \tau + g_A^{(0)}
%\right) + C^{(1)} \tau + C^{(0)},
%\end{equation}

{}From the solution~(\ref{solh}) for ${\cal A} - D$ and the gauge
conditions~(\ref{gauge1}) and (\ref{gauge2}), we can obtain the
expressions for ${\cal A}$ and $D$:
\begin{eqnarray}
{\cal A} &=& \frac{k}{2(k-1)} H - \sum_{A=1}^m
\frac{p_A+1}{(D-2) \Delta_A} \left( G_A + g_A^{(1)} \tau +
g_A^{(0)} \right) - \frac{B^{(1)} + C^{(1)}}{k-1} \tau -
\frac{B^{(0)} + C^{(0)}}{k-1},
\label{A} \\
D &=& \frac1{2(k-1)} H - \sum_{A=1}^m \frac{p_A+1}{(D-2)
\Delta_A} \left( G_A + g_A^{(1)} \tau + g_A^{(0)} \right) -
\frac{B^{(1)} + C^{(1)}}{k-1} \tau - \frac{B^{(0)} +
C^{(0)}}{k-1}, \label{D}
\end{eqnarray}
where the following constant parameters are introduced
\begin{equation}
C^{(1,0)} = \sum_{\rho=1}^p C_{\rho}^{(1,0)}.
\end{equation}

Finally we have to fulfill the last equation~(\ref{rr}). Using the
intersection rules~(\ref{intersection}) and the constraints
(\ref{ConBs}), this equation can be reduced to the following
constraint equations on the parameters:
\begin{equation}\label{Con}
\frac12 \sum_{A=1}^m \frac{\alpha_A^2 - (g_A^{(1)})^2}{\Delta_A} +
(B^{(1)})^2 + \sum_{\rho=1}^p (C_{\rho}^{(1)})^2 + \frac1{k-1} (
B^{(1)} + C^{(1)} )^2 + \frac12 (\phi^{(1)})^2 - \frac{k}{4(k-1)}
\beta^2 = 0.
\end{equation}

Let us count the number of free parameters. The total number of
parameters appearing in the solutions is $5m+2p+6$. It consists of
$2$ parameters of the function $H$ ($\beta, \tau_0$), $4m$
parameters in $G_A$ ($\alpha_A, \tau_A, g_A^{(0,1)}$), $2p+4$
parameters entering $B, C_{\rho}, \phi$ ($B^{(0,1)},
C_{\rho}^{(0,1)}, \phi^{(0,1)}$) and $m$ charge parameters $b_A$.
These have to satisfy $2m$ constraints (\ref{ConBs}) and one
constraint (\ref{Con}), which can fix, for example, $g_A^{(0,1)}$
and $\phi^{(1)}$. Thus the remaining number of independent
parameters is $3m+2p+5$. However, not all of these $3m+2p+5$
parameters are physical. The coordinate transformations allow us
to fix some of them: by rescaling $t, x_\rho$ one can absorb
$B^{(0)}$ and $C_{\rho}^{(0)}$. Also, since the system of
equations was autonomous, we are free to shift the coordinate
$\tau$ by a constant, so without loss of generality one can fix
$\tau_0 = 0$. Therefore we are free to fix $p+2$ and leave
$3m+p+3$ physical parameters, basically $3m$ of $b_A, \alpha_A$
and $\tau_A$ and $p+3$ of $B^{(1)}, C_{\rho}^{(1)}, \phi^{(0)}$
and $\beta$. Apparently, we also have a freedom of rescaling $r$,
but this is related to a change of the gauge function $\Lambda$.
So far we leave the constants in the expressions for the metric
functions unfixed for later convenience.

%%%%%%%%%%%%%%%%%%%%%%%%%%%%%%%%%%%%%%%%%%%%%%%%%%%%%%%%%%%%%%%%%%%%%%
\section{Fixing the constants}
%%%%%%%%%%%%%%%%%%%%%%%%%%%%%%%%%%%%%%%%%%%%%%%%%%%%%%%%%%%%%%%%%%%%%%
{}From now on we will consider only the topologically simple case
$\sigma=1$. We are interested in solutions (possibly) possessing
an event horizon and not plagued with naked singularities. The
simple, though incomplete, way to reveal the position of
singularities and to impose the regularity condition on the
horizon is to check the behavior of the Ricci scalar. Using the
Einstein equations one can find the following expression for the
Ricci scalar for the solution obtained:
\begin{equation}
\label{Ric} R = \frac{(k-1)^2}2 {\rm e}^{-2{\cal A}} \left(
\phi'^2 + \sum_{A=1}^m \frac{b_A^2 (D-2p_A-4)}{(k-1)^2(D-2)} {\rm
e}^{G_A} \right).
\end{equation}
Using the explicit form of $G_A$ in (\ref{ga1}), one can see that
the points $\tau = \tau_A$ are singular, unless some of them are
zero, in which case the singularity can be avoided by imposing
further conditions on the parameters. Other special points  are
$\tau \to \pm \infty$ (we assume the $\tau$-coordinate to vary on
the full real axis, and possibly to extend to the complex plane to
ensure the change of signs of the exponential terms like ${\rm
e}^{2B}$, see for details Ref.~\cite{Gal'tsov:2004kn}). These can
correspond to the horizons. At the horizons the metric coefficient
$g_{tt}$ must vanish. Using Eq.~(\ref{B}) one can see this can be
the case in the limits $\tau \to \pm \infty$ once suitable
inequalities on the parameters are imposed. Combining this with
the behavior of the Ricci scalar, we find, moreover, that one is
free to choose for the regular event horizon any of these two
limiting points, but then the other will be generically singular.
We choose $\tau \to -\infty$ as the horizon, then $\tau \to
\infty$ will be generically a null singularity, except for some
special choice of parameters.

Let us investigate the behavior of the metric functions at the
event horizon. Assuming without loss of generality $\beta \geq 0,
\alpha_A \geq 0$, (we will not consider the possibility of
imaginary values of these parameters which are also allowed by the
overall reality of the solution), we find that, as $\tau \to -
\infty$, the functions $G_A$ and $H$ become linear in $\tau$:
\begin{equation}\label{GHhor}
G_A \sim \alpha_A \, \tau, \quad H \sim \beta \, \tau.
\end{equation}
This ensures vanishing of the metric component ${\rm e}^{2B}$ at
the horizon.

An important further information can be extracted from the
constraint equation~(\ref{rr}). It is convenient to rewrite it as
follows:
\begin{equation}\label{conshor}
- {\cal A}'^2 + B'^2 + \sum_{\rho=1}^p C_\rho'^2 + k D'^2 +
\frac12 \phi'^2 = \sum_{A=1}^m \frac{b_A^2}2 {\rm e}^{G_A}- \sigma
k(k-1) {\rm e}^H.
\end{equation}
{}From the radial geodesic equation, it follows that ${\cal A}' =
B'$ at the horizon \cite{Gal'tsov:2004kn}, so the first two terms
on the left hand side cancel. The right hand side vanishes at the
horizon, so one is left with the sum of positive definite terms.
Therefore, all the derivatives $C_\rho',\, D',\, \phi'$ must
separately vanish in the limit $\tau \to -\infty$. More precisely,
the condition ${\cal A}' = B'$ gives
\begin{equation}\label{Ar}
\frac{k}{2(k-1)} \beta - \sum_{A=1}^m \frac{\alpha_A +
g_A^{(1)}}{\Delta_A} - \frac{B^{(1)} + C^{(1)}}{k-1} - B^{(1)} =
0.
\end{equation}
Moreover, using Eqs. (\ref{C}), (\ref{phi}) and (\ref{D}), we find
the following relations on the parameters involved, namely, from
Eq.~(\ref{C}) we obtain $p$ relations
\begin{equation}\label{Cr}
C_{\rho}^{(1)} = - \sum_{A=1}^m \frac{\delta_A^{(\rho)}}{D-2}
\frac{\alpha_A + g_A^{(1)}}{\Delta_A},
\end{equation}
{}from Eq. (\ref{phi})
\begin{equation}\label{phir}
\phi^{(1)} = \sum_{A=1}^m \epsilon_A a_A \frac{\alpha_A +
g_A^{(1)}}{\Delta_A},
\end{equation}
and from Eq. (\ref{D})
\begin{equation}\label{Dr}
\frac1{2(k-1)} \beta - \sum_{A=1}^m \frac{p_A+1}{D-2}
\frac{\alpha_A + g_A^{(1)}}{\Delta_A} - \frac{B^{(1)} +
C^{(1)}}{k-1} = 0.
\end{equation}
Consequently
\begin{equation}
C^{(1)} =  \sum_{A=1}^m \frac{D - 2 - k (p_A+1)}{D-2}
\frac{\alpha_A + g_A^{(1)}}{\Delta_A}.
\end{equation}

The above constraints (\ref{Ar})-(\ref{Dr}) are consistent with
(\ref{Con}). Combining (\ref{Ar}) and (\ref{Dr}), one can obtain
\begin{equation}
B^{(1)} = \frac12 \beta - \sum_{B=1}^m \frac{D-p_B-3}{D-2}
\frac{\alpha_B + g_B^{(1)}}{\Delta_B}.
\end{equation}
Furthermore, the constraints (\ref{ConBs}), after substituting
(\ref{Cr}) and (\ref{phir}), become
\begin{equation}
B^{(1)} = \frac12 \alpha_A - \sum_{B=1}^m \frac{D-p_B-3}{D-2}
\frac{\alpha_B + g_B^{(1)}}{\Delta_B}.
\end{equation}
Hence, all the parameters $\alpha_A$ must be equal:
\begin{equation}\label{alphaA}
\alpha_A = \beta.
\end{equation}
Then the metric functions read
\begin{eqnarray}
B &=& \sum_{A=1}^m \frac{D-p_A-3}{(D-2)\Delta_A} \left( G_A -
\beta \tau + g_A^{(0)} \right) + \frac12 \beta \tau + B^{(0)},
\\
C_{\rho} &=& \sum_{A=1}^m \frac{\delta^{(\rho)}_A}{(D-2) \Delta_A}
\left( G_A - \beta \tau + g_A^{(0)} \right) + C_\rho^{(0)},
\\
\phi &=& - \sum_{A=1}^m \frac{\epsilon_A a_A}{\Delta_A} \left( G_A
- \beta \tau + g_A^{(0)} \right) + \phi^{(0)},
\\
{\cal A} &=& \frac{k}{2(k-1)} H - \sum_{A=1}^m
\frac{p_A+1}{(D-2) \Delta_A} \left( G_A - \beta \tau +
g_A^{(0)} \right) - \frac{\beta}{2(k-1)} \tau - \frac{B^{(0)} +
C^{(0)}}{k-1},
\\
D &=& \frac1{2(k-1)} H - \sum_{A=1}^m \frac{p_A+1}{(D-2)
\Delta_A} \left( G_A - \beta \tau + g_A^{(0)} \right) -
\frac{\beta}{2(k-1)} \tau - \frac{B^{(0)} + C^{(0)}}{k-1}.
\end{eqnarray}
Though the parameters $\alpha_A$ are the same for all $A$, the
functions $G_A$ differ in position of singularities $\tau_A$

\subsection{Asymptotically flat solutions}
The asymptotic region is located at $\tau \to \tau_0$, and we have
fixed the translational freedom by choosing $\tau_0 = 0$. The
corresponding behavior of the function $H$ is
\begin{equation}
H \simeq \ln \frac1{\tau^2}.
\end{equation}
However, there are two different cases for $G_A$ depending on the
value of $\tau_A$. For the case $\tau_A \ne 0$, the asymptotic
value of $G_A$, for $\alpha_A = \beta$, is
\begin{equation}
G_A \simeq G_A^{(0)} := \ln \left[ \frac{\beta^2}{b_A^2 \Delta_A
\sinh^2(\frac{\beta}2 \tau_A)} \right].
\end{equation}

Therefore, asymptotically the solutions reduces to
\begin{eqnarray}
B &\simeq& \sum_{A=1}^m \frac{D-p_A-3}{(D-2)\Delta_A} \left(
G_A^{(0)} + g_A^{(0)} \right) + B^{(0)},
\\
C_{\rho} &\simeq& \sum_{A=1}^m \frac{\delta^{(\rho)}_A}{(D-2)
\Delta_A} \left( G_A^{(0)} + g_A^{(0)} \right) + C_{\rho}^{(0)},
\\
\phi &\simeq& - \sum_{A=1}^m \frac{\epsilon_A a_A}{\Delta_A}
\left( G_A^{(0)} + g_A^{(0)} \right) + \phi^{(0)}.
\end{eqnarray}
The quantities $B$ and $C_\rho$ are constants which can be set to
zero by rescaling of time and world-volume coordinates imposing
the conditions
\begin{eqnarray}
B^{(0)} &=& - \sum_{A=1}^m \frac{D-p_A-3}{(D-2)\Delta_A} \left(
G_A^{(0)} + g_A^{(0)} \right),
\\
C_{\rho}^{(0)} &=& -\sum_{A=1}^m \frac{\delta^{(\rho)}_A}{(D-2)
\Delta_A} \left( G_A^{(0)} + g_A^{(0)} \right).
\end{eqnarray}
For the dilaton, it is common to preserve the finite value
$\phi_\infty$ at infinity, and this can be ensured by the redefinition
\begin{equation}
\phi^{(0)} = \sum_{A=1}^m \frac{\epsilon_A a_A}{\Delta_A} \left(
G_A^{(0)} + g_A^{(0)} \right) + \phi^\infty.
\end{equation}
Then the relation (\ref{ConBs}) gives
\begin{equation}\label{ConphiA}
\phi^\infty  = - \frac{G_A^{(0)}}{\epsilon_A a_A}.
\end{equation}
This equation requires the quantities $G_A^{(0)}/\epsilon_A a_A$
to be identical for all $A$, which imposes a new set of relations
between $b_A$ and $\tau_A$. The Ricci scalar diverges at $\tau_A$,
so, to avoid naked singularities, one has to ensure that all
$\tau_A$ lie inside the event horizon for black branes. If there
is also an inner horizon, one can check that the real metrics
correspond to location of singularities $\tau_A$ either outside
the event horizon, or inside the inner horizon. The first
possibility should be ruled out. The notable exception constitutes
the case of $\tau_A = 0$. In this case there is no singularity,
and this point can be interpreted as spatial infinity. The
physical nature of this will be clearer later when we express the
solution in the Schwarzschild-type coordinate.

\subsection{Asymptotically LDB branes}
For the case $\tau_A = 0$, each $G_A$ has the same asymptotic
behavior as $H$:
\begin{equation}
G_A \simeq \ln \frac1{\tau^2}.
\end{equation}
In such case, we have
\begin{eqnarray}
B &\simeq& \sum_{A=1}^m \frac{D-p_A-3}{(D-2)\Delta_A} \,
\ln \frac1{\tau^2},
\\
C_{\rho} &\simeq& \sum_{A=1}^m \frac{\delta^{(\rho)}_A}{(D-2)
\Delta_A} \, \ln \frac1{\tau^2},
\\
\phi &\simeq& - \sum_{A=1}^m \frac{\epsilon_A a_A}{\Delta_A} \,
\ln \frac1{\tau^2}.
\end{eqnarray}
This solution is not asymptotically flat, but asymptotically
approaching the LDB.

%%%%%%%%%%%%%%%%%%%%%%%%%%%%%%%%%%%%%%%%%%%%%%%%%%%%%%%%%%%%%%%%%%%%%%
\section{Schwarzschild-type coordinates}
%%%%%%%%%%%%%%%%%%%%%%%%%%%%%%%%%%%%%%%%%%%%%%%%%%%%%%%%%%%%%%%%%%%%%%
To present our solution in a more familiar form, we change the
coordinates similarly to \cite{Gal'tsov:2004kn}. We map the
horizon $\tau = -\infty$ to $r = r_H$,
\begin{equation}
r_H = \mu^{\frac1{k-1}}, \qquad \mu = \frac{\beta}{k-1}.
\end{equation}
and the internal horizon $\tau = \infty$ to $r = 0$ by choosing
the gauge function
\begin{equation}\label{La}
\Lambda = \ln ( r^k f ), \qquad \mbox{such that} \qquad d\tau =
\frac{dr}{r^k f},
\end{equation}
where
\begin{equation}\label{fpm}
f = 1 - \frac{\mu}{r^{k-1}}.
\end{equation}
This corresponds to the coordinate transformation
\begin{equation}\label{tau-r}
\tau = \frac1{(k-1)\mu} \ln f.
\end{equation}
Thus the region outside the event horizon $r > r_H$ corresponds to
the half axis $(-\infty,\; 0)$ of $\tau$. Then we will have
\begin{equation}
H = \ln \left( r^{2(k-1)} f  \right).
\end{equation}

\subsection{Asymptotically flat solutions}

Introduce a new parameter $q_A$ instead of $\tau_A$ by
\begin{equation}\label{qA}
q_A = \frac{\mu}{{\rm e}^{\mu(k-1) \tau_A} - 1}.
\end{equation}
As we already noted, to ensure the absence of naked singularities,
all $\tau_A$ should be taken non-negative. Here we assume that
$\tau_A$ are strictly positive, so this definition leads to finite
$q_A$ (the case $\tau_A=0$ will be considered below). The function
$G_A$ then reads
\begin{equation}
G_A = \ln \left( \frac{4(k-1)^2 (\mu + q_A ) q_A}{\Delta_A b_A^2}
\frac{f }{h_A^2} \right),
\end{equation}
where
\begin{equation}
h_A = 1 + \frac{q_A}{r^{k-1}},
\end{equation}
are the harmonic functions.

In terms of the new coordinates, the metric
components~(\ref{metric}) read
\begin{eqnarray}
B &=& - 2 \sum_{A=1}^m \frac{D-p_A-3}{(D-2)\Delta_A} \ln h_A +
\frac12 \ln f,
\\
C_{\rho} &=& -2 \sum_{A=1}^m \frac{\delta^{(\rho)}_A}{(D-2)
\Delta_A} \ln h_A,
\\
\phi &=& 2 \sum_{A=1}^m \frac{\epsilon_A a_A}{\Delta_A} \ln h_A +
\phi^\infty,
\\
A &=& - \frac12 \ln f + 2 \sum_{A=1}^m \frac{p_A+1}{(D-2)
\Delta_A} \ln h_A,
\\
D &=& \ln r + 2 \sum_{A=1}^m \frac{p_A+1}{(D-2) \Delta_A} \ln h_A,
\end{eqnarray}
with a constraint following from (\ref{ConphiA}):
\begin{equation}
b_A^2 = {\rm e}^{\epsilon_A a_A \phi^\infty} \frac{4(k-1)^2 (\mu
+ q_A) q_A}{\Delta_A}.
\end{equation}
For the electric branes, the solution for the form field is
\begin{equation}
f^{\rm elec}_A = b_A {\rm e}^{- a_A \phi^\infty} \frac1{r^k
h_A^2}.
\end{equation}
This solution corresponds to the one given in \cite{Ohta:1997gw}. Free
parameters of the solution are $\mu, q_A$ and $\phi_\infty$.

\subsection{Asymptotically LDB branes}
If $\tau_A = 0$, $G_A$ takes the form
\begin{equation}
G_A = \ln \left( \frac{4(k-1)^2}{\Delta_A b_A^2} r^{2(k-1)} f
\right).
\end{equation}
Then the solution is, with the choice of parameters $B^{(0)}=0$
and $C_\rho^{(0)}=0$,
\begin{eqnarray}
B &=& 2 \sum_{A=1}^m \frac{D-p_A-3}{(D-2)\Delta_A} \ln
\frac{r^{k-1}}{q_A} + \frac12 \ln f,
\\
C_{\rho} &=& 2 \sum_{A=1}^m \frac{\delta^{(\rho)}_A}{(D-2)
\Delta_A} \ln \frac{r^{k-1}}{q_A},
\\
\phi &=& - 2 \sum_{A=1}^m \frac{\epsilon_A a_A}{\Delta_A} \ln
\frac{r^{k-1}}{q_A} + \phi^\infty,
\\
A &=& - \frac12 \ln f - 2 \sum_{A=1}^m \frac{p_A+1}{(D-2)
\Delta_A} \ln \frac{r^{k-1}}{q_A},
\\
D &=& \ln r - 2 \sum_{A=1}^m \frac{p_A+1}{(D-2) \Delta_A} \ln
\frac{r^{k-1}}{q_A},
\end{eqnarray}
where the parameters $q_A$ are defined as
\begin{equation}
q_A^{-2} = \frac{4(k-1)^2}{\Delta_A b_A^2} \, {\rm e}^{g_A^{(0)}}.
\end{equation}
Our previous form strength parameter $b_A$ is related to $ q_A$
via
\begin{equation}
b_A^2 = {\rm e}^{\epsilon_A a_A \phi^\infty} \frac{4(k-1)^2 \,
q_A^2}{\Delta_A}.
\end{equation}
The solution for the electric form field is
\begin{equation}
f^{\rm elec}_A = b_A {\rm e}^{- a_A \phi^\infty}
\frac{r^{k-2}}{q_A^2}.
\end{equation}

This second possibility for $\tau_A$ leads to an intersecting
generalization of the solution of Ref. \cite{Clement:2004ii}. The
BPS limit of these solutions corresponds to $\mu=0$. It can be
recognized that in this case one deals with a solution known
earlier as the hear-horizon limit of the BPS intersecting branes.
This corresponds to the usual rule of omitting constants in the
harmonic functions describing the metric. Here we reproduce these
metrics as a particular case of the general supergravity solution.
Our result is, however, not just to reproduce them by another
technique, but to prove that no other solutions without naked
singularities exist within the class of metrics considered, which
is fixed by their isometries, up to the the mixed intersections
below.

The black versions of these solutions with $\mu\neq 0$ should
describe the thermal phase of QFT's in the Domain-Wall/QFT correspondence
associated to their BPS limit.

\subsection{Mixed intersections}

The mixed intersecting configurations, namely the black AF
$p$-branes with the asymptotically LDB ones, can be obtained
simply by picking up the corresponding term in the summation of
$A, B, C_\rho, D$ and $\phi$ depending on whether $p_A$ is AF
brane or LDB. Again, the BPS limit of these solutions was found
previously through the near-horizon considerations. It was noted
that one can drop the constant term in the harmonic functions
describing the system of intersecting branes not in all harmonic
functions at the same time, but in some of them \cite{Bo97}. This
corresponds to the BPS limit of our mixed intersection. To our
knowledge, the black version of these solutions is new.

%%%%%%%%%%%%%%%%%%%%%%%%%%%%%%%%%%%%%%%%%%%%%%%%%%%%%%%%%%%%%%%%%%%%%%
\section{Two intersecting branes }
%%%%%%%%%%%%%%%%%%%%%%%%%%%%%%%%%%%%%%%%%%%%%%%%%%%%%%%%%%%%%%%%%%%%%%
We give the examples of two intersecting branes. According to the
intersection rules, we can have solutions listed below:
\begin{eqnarray}
\mbox{IIA} &:& 2 \perp 2(0), \quad 4 \perp 4(2), \quad 6 \perp
6(4), \quad 0 \perp 4(0), \quad 2 \perp 4(1), \quad 2 \perp 6(2),
\quad 4 \perp 6(3), \quad 4 \perp 8(4), \quad 6 \perp 8(5),
\nonumber\\
\mbox{IIB} &:& 3 \perp 3(1), \quad 5 \perp 5(3), \quad 7 \perp
7(5), \quad 1 \perp 3(0), \quad 1 \perp 5(1), \quad 3 \perp 5(2),
\quad 3 \perp 7(3), \quad 5 \perp 7(4), \quad 5 \perp 9(5),
\nonumber\\
\mbox{M} &:& {\rm M}2 \perp {\rm M}2(0), \quad {\rm M}2 \perp {\rm
M}5(1), \quad {\rm M}5 \perp {\rm M}5(3),
\nonumber\\
\mbox{NS} &:& {\rm NS}1 \perp {\rm NS}5(1), \quad {\rm NS}5 \perp
{\rm NS}5(3), \qquad {\rm D}p \perp {\rm NS}1(0), \, 0 \le p \le
8, \qquad {\rm D}p \perp {\rm NS}5(p-1), \, 1 \le p \le 6,
\end{eqnarray}
where those branes indicated only by numbers are D-branes, and the
number in the parenthesis denotes the dimensions of the
overlapping coordinates. We give explicit solutions for M and IIB
theories in the following.

\subsection{M2-M2 configurations}

Let us consider the general solution for two M2-branes as an
explicit example. In this case, the parameters have the values:
$D=11, k = 5, n_A = 4, a_A = 0, p_A+1 = 3, D-p_A-3 = 6$ and
$\Delta_A = 4$. The AF solutions have the metric
\begin{equation}
ds^2 = h_1^{\frac13} h_2^{\frac13} \left[ - f h_1^{-1} h_2^{-1}
dt^2 +  h_1^{-1} (dx_1^2 + dx_2^2) +  h_2^{-1} (dx_3^2 + dx_4^2) +
f^{-1} dr^2 +  r^2 d\Omega_5^2 \right],
\end{equation}
where $h_A = 1 + q_A/r^4$, and the form fields are
\begin{equation}
f_A^{\rm elec} = \frac{b_A}{r^5 h_A^2}, \qquad b_A^2 = 16 \, ( \mu
+ q_A ) q_A.
\end{equation}
We recover the canonical intersecting two M2-branes.

For asymptotically LDB case, the solution we have derived
corresponds to omitting the constant in the harmonic functions
$h_1$ and $h_2$. It is
\begin{equation}
ds^2 = \left( \frac{q_1}{r^4} \right)^{\frac13} \left(
\frac{q_2}{r^4} \right)^{\frac13} \left[ - f \left(
\frac{q_1}{r^4} \right)^{-1} \left( \frac{q_2}{r^4} \right)^{-1}
dt^2 + \left( \frac{q_1}{r^4} \right)^{-1} (dx_1^2 + dx_2^2) +
\left( \frac{q_2}{r^4} \right)^{-1} (dx_3^2 + dx_4^2) + f^{-1}
dr^2 + r^2 d\Omega_5^2 \right],
\end{equation}
and
\begin{equation}
f_A^{\rm elec} = \frac{b_A r^3}{q_A^2}, \qquad b_A^2 = 16 \,
q_A^2.
\end{equation}

The mixed intersecting configurations of two M2-branes is given by
\begin{equation}
ds^2 = h_1^{\frac13} \left( \frac{q_2}{r^4} \right)^{\frac13}
\left[ - f h_1^{-1} \left( \frac{q_2}{r^4} \right)^{-1} dt^2 +
h_1^{-1} (dx_1^2 + dx_2^2) + \left( \frac{q_2}{r^4} \right)^{-1}
(dx_3^2 + dx_4^2) + f^{-1} dr^2 + r^2 d\Omega_5^2 \right],
\end{equation}
where
\begin{equation}
f_1^{\rm elec} = \frac{b_1}{r^5 h_1^2}, \quad b_1^2 = 16 ( \mu +
q_1 ) q_1 , \qquad f_2^{\rm elec} = \frac{b_2 r^3}{q_2^2}, \quad
b_2^2 = 16 \, q_2^2.
\end{equation}

\subsection{D1-D5 configurations}

The other interesting case is the D1-D5 system. In this case, the
parameters have the following values: $D=10, k = 3, n_A = 3, a_A =
1, \epsilon_1 = 1, \epsilon_2 = -1, p_1 + 1 = 2 = D - p_2 - 3, p_2
+ 1 = 6 = D - p_1 - 3$ and $\Delta_A = 4$. The AF solutions have
the metric
\begin{equation}
ds^2 = h_1^{\frac14} h_2^{\frac34} \left[ h_1^{-1} h_2^{-1} ( - f
dt^2 + dx_1^2 ) +  h_2^{-1} (dx_2^2 + \cdots + dx_5^2) + f^{-1}
dr^2 +  r^2 d\Omega_3^2 \right],
\end{equation}
and the dilaton and the form fields
\begin{equation}
{\rm e}^{2\phi} = {\rm e}^{2 \phi^\infty} \frac{h_1}{h_2}, \qquad
f_1^{\rm elec} = {\rm e}^{- \phi^\infty} \frac{b_1}{r^3 h_1^2},
\quad  b_A^2 = {\rm e}^{\epsilon_A \phi^\infty} 4 \, ( \mu + q_A )
q_A,
\end{equation}
where $h_A = 1 + q_A/r^2$. We recover the canonical intersecting
D1-D5-branes in the Einstein frame.

Again, for LDB,  our solution corresponds to omitting the constant
in the harmonic functions $h_1$ and $h_2$. It is
\begin{equation}
ds^2 = \left( \frac{q_1}{r^2} \right)^{\frac14} \left(
\frac{q_2}{r^2} \right)^{\frac34} \left[ \left( \frac{q_1}{r^2}
\right)^{-1} \left( \frac{q_2}{r^2} \right)^{-1} ( - f dt^2 +
dx_1^2 ) + \left( \frac{q_2}{r^2} \right)^{-1} (dx_2^2 + \cdots
dx_5^2) + f^{-1} dr^2 + r^2 d\Omega_3^2 \right],
\end{equation}
and
\begin{equation}
{\rm e}^{2\phi} = {\rm e}^{2 \phi^\infty} \frac{q_1}{q_2}, \qquad
f_1^{\rm elec} = {\rm e}^{- \phi^\infty} \frac{b_1 r}{q_1^2},
\quad b_A^2 = {\rm e}^{\epsilon_A \phi^\infty} 4 \, q_A^2.
\end{equation}

There are also two mixed intersecting configurations of
D1-D5-branes. One is given by
\begin{equation}
ds^2 = h_1^{\frac14} \left( \frac{q_2}{r^2} \right)^{\frac34}
\left[ h_1^{-1} \left( \frac{q_2}{r^2} \right)^{-1} (- f dt^2 +
dx_1^2) + \left( \frac{q_2}{r^2} \right)^{-1} (dx_2^2 + \cdots +
dx_5^2) + f^{-1} dr^2 + r^2 d\Omega_3^2 \right],
\end{equation}
where
\begin{equation}
{\rm e}^{2\phi} = {\rm e}^{2 \phi^\infty} \frac{r^2 h_1}{q_2},
\qquad f_1^{\rm elec} = {\rm e}^{- \phi^\infty} \frac{b_1}{r^3
h_1^2}, \quad b_1^2 = {\rm e}^{\phi^\infty} 4 ( \mu + q_1 ) q_1 ,
\qquad b_2^2 = {\rm e}^{- \phi^\infty} 4 \, q_2^2,
\end{equation}
and the other is
\begin{equation}
ds^2 =  \left( \frac{q_1}{r^2} \right)^{\frac14} h_2^{\frac34}
\left[ \left( \frac{q_1}{r^2} \right)^{-1} h_2^{-1} (- f dt^2 +
dx_1^2) + h_2^{-1} (dx_2^2 + \cdots + dx_5^2) + f^{-1} dr^2 + r^2
d\Omega_3^2 \right],
\end{equation}
where
\begin{equation}
{\rm e}^{2\phi} = {\rm e}^{2 \phi^\infty} \frac{q_1}{r^2 h_2},
\qquad f_1^{\rm elec} = {\rm e}^{- \phi^\infty} \frac{b_1
r}{q_1^2}, \quad b_1^2 = {\rm e}^{\phi^\infty} 4 \, q_1^2, \qquad
b_2^2 = {\rm e}^{- \phi^\infty} 4 \, ( \mu + q_2 ) q_2.
\end{equation}

%%%%%%%%%%%%%%%%%%%%%%%%%%%%%%%%%%%%%%%%%%%%%%%%%%%%%%%%%%%%%%%%%%%%%%
\section{Conclusions}
%%%%%%%%%%%%%%%%%%%%%%%%%%%%%%%%%%%%%%%%%%%%%%%%%%%%%%%%%%%%%%%%%%%%%%
In this paper we obtained the general solution for the non-BPS
intersecting supergravity $p$-branes delocalized in relative
transverse dimensions. This was done by fully integrating the
Einstein equations with a suitable ansatz for the metric, the
antisymmetric forms and the dilaton. Our solution differs from
earlier ones by a larger number of free parameters, which is
maximal in the present case. This, in particular, opens a way to
construct individual brane solutions which are either
asymptotically flat, or asymptotically approaching the LDB. For
intersecting branes, imposing the conditions on parameters which
ensure the absence of naked singularities, we found that the
resulting configuration is an intersection of some AF branes with
some asymptotically LDB branes. The intersection rules remain the
same as previously known. In the case when all branes are AF, we
recover the solution of \cite{Ohta:1997gw}. When all individual
branes are asymptotically LDB, we obtain the solution which
corresponds to omitting constant terms in the harmonic functions
involved (the overall solution being black). Apart from these two
extremes, there are intersections of a number of AF branes with a
number of LDB branes within the intersection rules.

The linear dilaton backgrounds were extensively used in the
context of DW/QFT (NS/LST) dualities \cite{Boonstra:1998mp,
Behrndt:1999mk,Aharony:1998ub, Aharony:1999ks}. The LDB solutions
endowed with the event horizon describe the thermal phase of the
QFT (LST) associated with the linear dilaton background
\cite{HaOb,BeRo,BoRo}. Although this was not our intention here,
our new intersecting solutions with LDB asymptotics may serve a
basis for further studies along these directions.

The BPS versions of our solutions involving intersections with LDB
branes correspond to partial omission of the constant terms in the
harmonic functions involved. It has to be emphasized that our
results are based on the complete integration of the Einstein
equations for metric of chosen symmetry, so there are no other
other solutions within this class which are free from naked
singularities. Various suggestions for $p$-brane solutions with
extra parameters which obey the same metric ansatz, but do not
reduce to the standard BPS or black branes, generically suffer
from naked singularities.

\begin{acknowledgments}
CMC would like to thank Koryu Kyokai grant for supporting his
visit to Japan and the hospitality of Yukawa Institute of
Theoretical Physics and Osaka University during this work was
proceeding. The work of CMC was supported by the NSC grant
93-2112-M-008-021 and 94-2119-M-002-001.
The work of NO was supported in part by the Grant-in-Aid for
Scientific Research Fund of the JSPS No. 16540250,
and that of DG was supported in part by the RFBR grant 02-04-16949.
\end{acknowledgments}

%%%%%%%%%%%%%%%%%%%%%%%%%%%%%%%%%%%%%%%%%%%%%%%%%%%%%%%%%%%%%%%%%%%%%%
%\begin{references}

\end{document}